\documentclass{ametsocV6.1}
\usepackage{tikz}

\definecolor{clrA}{rgb}{0.8500,0.3250,0.0980} 
\definecolor{clrB}{rgb}{0.3010,0.7450,0.9330} 
\definecolor{clrC}{rgb}{0.9290,0.6940,0.1250} 

\usetikzlibrary{shapes.geometric}
\NewDocumentCommand{\symHexFill}{O{clrC} O{3.1mm}}{%
  \tikz[baseline=-0.6ex]{%
    \node[
      star, star points=6, star point ratio=2.0, 
      draw=black, fill=#1,
      inner sep=0pt,
      minimum size=#2,
      line width=0.28mm
    ] {};
  }%
}

\NewDocumentCommand{\symPentaFillC}{O{clrC} O{3.1mm}}{%
  \tikz[baseline=-0.6ex]{%
    \node[
      star, star points=5, star point ratio=2.1,
      draw=black, fill=#1, inner sep=0pt,
      minimum size=#2, line width=0.28mm
    ] {};
  }%
}

\NewDocumentCommand{\symPentaFillB}{O{clrB} O{3.1mm}}{%
  \tikz[baseline=-0.6ex]{%
    \node[
      star, star points=5, star point ratio=2.1,
      draw=black, fill=#1, inner sep=0pt,
      minimum size=#2, line width=0.28mm
    ] {};
  }%
}

\title{Surface Wave-Aerodynamic Roughness Length Model\\
for Air-Sea Interactions}

\authors{Manuel Ayala,\aff{a}\correspondingauthor{mayala5@jhu.edu} 
Dennice F. Gayme,\aff{a} 
Charles Meneveau,\aff{a} 
}

\affiliation{\aff{a}{Department of Mechanical Engineering, Johns Hopkins University, Baltimore, USA}
}

\abstract{A new model to evaluate the equivalent hydrodynamic length  or surface roughness, $z_0$, of ocean waves is developed and tested. The proposed Surface Wave-Aerodynamic Roughness Length (SWARL) model requires maps of the wave surface height at consecutive times and  the air flow characteristic Reynolds number as inputs. Pressure drag is accounted for by approximating the  relative velocity in a frame moving with the local wave phase-speed assuming  ideal inviscid ramp flow \citep{ayala2024}. Drag from viscous and unresolved ripples is modeled using the standard equilibrium model. The SWARL model is tested using over 300 datasets for monochromatic and broad-spectrum wave surfaces. The model-predicted $z_0$ and drag coefficients are compared to measured values, as well as commonly used wave parametrization methods found in the literature. For datasets with well-characterized surfaces, the proposed model shows significantly better agreement with data compared to prior models. For data  that did not include a full characterization of the wave fields (typically field data), the model yields predictions with accuracy similar to prior models. Results highlight that including detailed flow physics and extensive wave-field characterization in the modeling of $z_0$ can provide significant improvements in roughness-length based modeling of air-sea interactions.}

\begin{document}

\maketitle

%
%
%
\statement

This paper introduces a new method to predict the surface drag force of wind over ocean waves, a key parameter for weather forecasting, climate modeling, and offshore engineering. Drag is typically quantified by a roughness length (a measure of the resistance to the wind at the wind-wave interface due to moving surface waves), which conventional methods estimate using empirical formulas or assumptions that fail to capture the dynamics accurately. Our model predicts the roughness length based on geometric knowledge about the moving surface. It performs well on both simple and complex wave shapes, showing significantly better agreement with data than traditional approaches. By improving drag predictions, this method can enhance simulations of hurricane prediction, climate modeling, and offshore wind farm design.

\section{Introduction}
\label{sec:intro}
The interaction between wind and waves represents one of nature's most intricate, multiscale phenomena, occurring at the crucial interface between the atmosphere and the ocean \citep{young1999}. Accurate prediction of the associated fluxes (of momentum, heat, water, air) is essential for advancing weather forecasting, climate studies \citep{cronin_review_2019}, and enhancing offshore  wind farm design and construction, especially as the demand for sustainable energy sources such as offshore wind farms, increases \citep{veers2023}. 
However, developing accurate and practical parameterizations of air-sea surface fluxes remains challenging due to the inherent complexity of marine atmospheric boundary layer (MABL) turbulence.  Similar to static surface roughness, moving waves play a critical role in shaping the momentum balance between turbulent air flow and the underlying surface~\citep{sullivan_annual2010, chung_review_2021}.
However, unlike static roughness elements that are fixed and interact with the flow only through form and viscous drag due to the relative velocity between air and the surface, wave fields move, and are characterized by local phase velocity and orbital velocities that affect the air flow above. Nonetheless, flow over both surfaces often result in logarithmic (or Monin-Obukov similarty theory based) profiles of the mean air velocity, at sufficiently high Reynolds numbers. Therefore, the averaged wave effects can be similarly represented with an equivalent hydrodynamic length, $z_0$ \citep{Deskos2021}. Determining  this quantity accurately based on information about the surface and without having to perform costly experiments or eddy and surface-resolving numerical simulations remains an open challenge.

In the case of static surface roughness, traditional methods to determine $z_0$ utilize topographical parameters that depend solely on the geometry of the surface \citep{chung_review_2021}. Approaches for estimating equivalent roughness lengths   \textit{a priori} have been developed with some success using empirically fit models \citep{flack_chung_2022}, machine learning approaches \citep{AghaeiJouybari2021}, and more recently, a fluid mechanics-based geometric parameter called the wind-shade factor \citep{meneveau2024}. On the other hand, due to the inherently time-varying nature of waves, most of the ideas and methods that work for static surfaces are not applicable. 

State-of-the-art approaches  for determining equivalent surface roughness for surface waves similarly rely heavily on empirical models with parameters obtained via fits to available data \citep{Deskos2021}. Using dimensional analysis, \citet{charnock} proposed the earliest and best-known roughness parametrization for ocean waves as:
\begin{equation}
    z_0/H_s = \alpha_{\text ch}u_*^2/(g H_s),
    \label{charnock}
\end{equation}
where $z_0$ is the equivalent hydrodynamic length (roughness length), $H_s$ is the significant wave height, $u_*$ is the friction velocity, and $g$ is the gravitational acceleration. The Charnock parameter $\alpha_{\text{ch}}$  was originally treated as a constant, though subsequent studies have demonstrated that it varies significantly with sea state and environmental conditions  \citep{Deskos2021}. One of the most widely used surface flux parameterizations is the Coupled Ocean–Atmosphere Response Experiment (COARE) 3.5 bulk flux algorithm \citep{edson2013}.  The COARE 3.5 algorithm provides two parameterizations for $\alpha_{\text{ch}}$, one based on wave age:
\begin{equation}
    \alpha_{\text ch} = 0.114(u_*/c_p)^{0.622},
    \label{coare1}
\end{equation}
and another based on wave steepness:
\begin{equation}
    \alpha_{\text ch} = 0.091\,H_sk_p,
    \label{coare2}
\end{equation}
where $c_p$ and $k_p$ are the wave-field's peak wave phase speed and wave-number, respectively. Wave forecasting models such as WAve Model (WAM), Simulating WAves Nearshore (SWAN), and WaveWatch III (WW3) adopt yet another approach for estimating the Charnock parameter. In these models, it is calculated as:
\begin{equation}
    \alpha_{\text ch} = \frac{\hat{\alpha}}{\sqrt{1-\frac{\tau_w}{\tau}}},
\end{equation}
where, $\hat{\alpha}$ is typically taken between 0.01 and 0.0185 \citep{ecmwf_wave_model_2016, swan_tech_doc_2024}, $\tau = \rho_a \, u_*^2$ is the total stress at the interface between wind and waves,  where $\rho_a $ is the density of the air. $\tau_w$ is the wave-induced stress calculated as:
\begin{equation}
\tau_w = \rho_a \int\int \omega^2 \, \gamma\, E(|k|,\theta)\, d\theta |k| dk,
\end{equation}
where $\omega$ is the angular frequency, $E(|k|,\theta)$ is the two-dimensional  wave energy spectrum and $\gamma$ is the growth rate parameter, which is commonly taken as  
\begin{equation}
\gamma = 0.25 \left(  \frac{u_*}{c} \cos{\theta_w} -1 \right)
\end{equation}
in WAM \citep{wamdi_group_1988} and SWAN \citep{swan_tech_doc_2024}. Many other studies have proposed parameterizations of roughness based on wave age, defined as $c_p/u_*$.
These models generally take the form:
\begin{equation}
    z_0/H_s = A(u_*/c_p)^B,
    \label{zo_waveage}
\end{equation}
where, $A$ and $B$ are two empirically-determined parameters from fitting to data. \citet{Drennan2003} proposed $A = 3.35$ and $B= 3.4$ while \citet{Donelan1990} proposed $A = 0.46$ and $B= 2.53$. 
To account for wind-wave misalignment, \citet{porchetta2019} introduced a directional dependence in the coefficients and proposed $A = 20\cos{(0.45\,\theta_w)}$ and $B=3.82 \cos{(-0.32\,\theta_w)}$, where $\theta_w$ is the angle between wind and wave direction. An alternative roughness parameterization based on wave steepness was proposed by \citet{taylor_yelland}, leading to the expression:
\begin{equation}
    z_0/H_s = 1200(H_s/\lambda_p)^{3.4},
    \label{TY}
\end{equation}
where $\lambda_p$ is the wavelength of the peak wave.These surface roughness parameterizations form a foundational component of many atmospheric and oceanographic modeling systems. For a more comprehensive discussion of these and other models, we refer to \cite{zhao2024, linsheng2020, Deskos2021}.

Surface roughness models are routinely used to estimate momentum and scalar fluxes in a wide range of applications. For example, mesoscale simulations of offshore wind farms within the Weather Research and Forecasting (WRF) framework \citep{jimenez_wrf_2015}, as well as hurricane modeling studies \citep{davis_hurricanes_wrf_2008} rely on roughness length parameterizations to represent air-sea interactions. Global climate models also utilize this type of approach to account for surface-atmosphere exchanges \citep{Couvelard2020}. At finer scales, Large Eddy Simulations (LES) of marine boundary layers using WRF-LES \citep{munoz_esparza_2014, NING2023105592} and offshore wind turbine flows \citep{Johlas_2020, YANG2022124674} incorporate $z_0$-based models to capture surface heterogeneity and wave effects. While these models have enabled substantial advances, they often lack robustness across a wide range of sea states and wave conditions \citep{cronin_review_2019}, prompting ongoing efforts to develop more universal and physics-based parameterizations that reduce reliance on empirical formulae and parameter tuning.

In this article, we describe a new method to determine $z_0$ for a given moving wave-field. The proposed Surface Wave-Aerodynamic Roughness Length (SWARL) model   requires  maps of a wave field's spatial distribution of surface heights over a representative horizontal extent at two successive times, and the relevant Reynolds number as inputs. A scalar parameter, $\Lambda$, is then evaluated numerically (or analytically under simplifying assumptions) as a surface average of geometric surface properties. This  $\Lambda$ includes information regarding form drag due to waves and wave history effects (i.e. wave phase-velocity) based on the local surface inclination with respect to the relative air-velocity. An efficient iterative procedure is used to  capture dependence on wave age and Reynolds number.  Additional drag from viscous and small-scale ripples is modeled using standard equilibrium surface layer modeling concepts that can also be  included in the evaluation of the  $\Lambda$ parameter. The proposed model is validated against a large number of data sets, as well as existing models for monochromatic and multiscale waves.

\section{The Surface Wave Aerodynamic Roughness Length (SWARL) model}
\label{sec:wave_drag_model}

We begin by expressing the momentum exchange (total drag force in the $i^{{th}}$ direction on planform area $A$) as an integral of a surface stress,   $\tau^{\text{w}}_{i3}$ where $i=1,2$ are the wind streamwise $x$-direction and transverse $y$ directions, and 3 is the vertical $z$ direction). The force in the $x$ direction (aligned with the air-flow) is written as   $ \iint_A \, \tau^{\text{w}}_{13}\,\, dx dy = u_*^2 A$, where $u_*$ is the friction velocity. The wall stress $\tau^{\text{w}}_{13}=\tau^{\text{w-p}}_{13}+\tau^{\text{w}-u}_{13}$ consists of two contributions: pressure (form drag) $\tau^{\text{w-p}}_{13}$, and drag from unresolved effects, like viscous and small-scale roughness (smaller than the resolved ones causing pressure stress) $\tau^{\text{w}-u}_{13}$.

\begin{figure}[h]
 \centerline{\includegraphics[width=30pc]{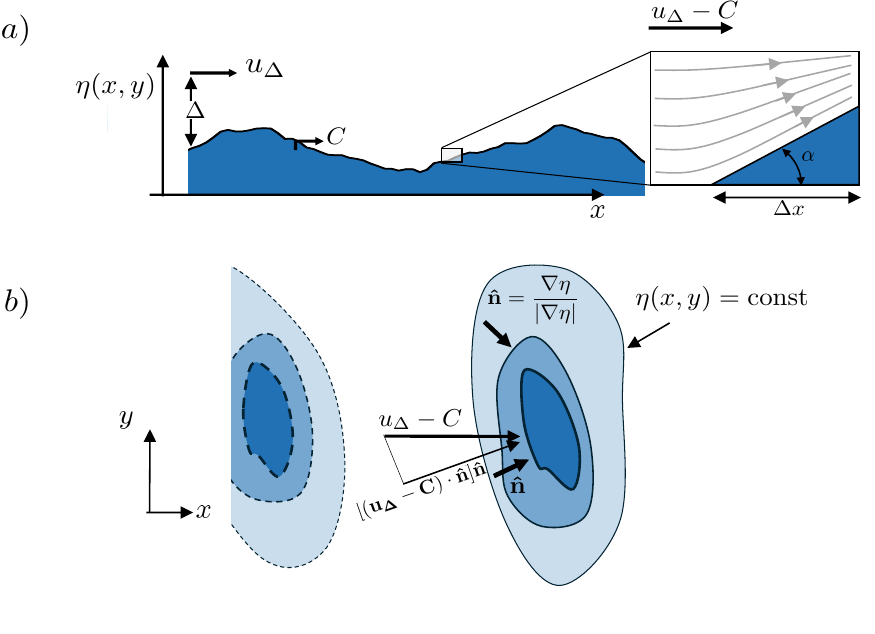}}
  \caption{a) Sketch of an instantaneous wave-field   surface distribution and potential flow over a ramp at angle $\alpha$, assumed to represent the local flow over the surface over a spatial extent $\Delta x$. In the sketch, it is assumed that the wave-field is 2D, with local surface normal ${\bf{\hat{n}}} = {{\bf{{\nabla}}\eta} }/{|{\bf{{\nabla}}\eta}|}$ in the same direction ($x)$ as the relative velocity. b) Top-view sketch of wave-field isosurface contours, when surface elevation also varies in the transverse ($y$) direction. The sketch shows the local normal vector ${\bf{\hat{n}}}$, the incoming relative velocity $(u_\Delta-C)$ and the incoming velocity normal to the surface which has a value equal to $({\bf{u_\Delta-C}})\cdot{\bf{\hat{n}}}$ and is in the $\bf{\hat{n}}$ direction.} 
  \label{fig0}
\end{figure}

The pressure stress is expressed using the wall stress model for moving surfaces introduced in \citet{ayala2024}. The model assumes that in the frame of the wave's local phase speed ${\bf C}$ with incoming relative velocity ${\bf u}_\Delta-{\bf C}$ (where ${\bf u}_\Delta$ is the horizontal air velocity at some reference height $\Delta$ above the wave mean elevation height), the local flow can be represented as potential flow over a ramp with slope angle $\alpha(x,y)$. Figure \ref{fig0}(a) shows a schematic representation of  potential flow over a ramp, illustrating the basic modeling assumption.

The average pressure over such a ramp is proportional to the relative velocity squared multiplied by $\alpha/(\pi+\alpha)$ and the pressure contribution (form drag) to the modeled stress can be written as 
\begin{equation}
\tau^{\text{w-p}}_{i3} = \frac{\alpha}{\pi\,+\,\alpha}\, |(\boldsymbol{u_\Delta-C}) \cdot \boldsymbol{\hat{n}}|^2\,|\boldsymbol{\nabla} \eta|\,\, \text{H} \left[ (\boldsymbol{u_\Delta}-\boldsymbol{C})\cdot \boldsymbol{\nabla} \eta \right] \, \hat{n}_i\qquad i = 1,2.
\label{eq:mosd_wpm}
\end{equation} 
Here $\eta(x,y,t)$ is the  known surface height distribution as function of horizontal positions $(x,y)=(x_1,x_2)$ and time $t$, $\alpha(x,y) = \arctan{|\boldsymbol{\nabla} \eta|}$. The pressure rise is due to the relative velocity normal to the surface. Therefore only the normal projection of the relative velocity $({\bf u}_\Delta-{\bf C})\cdot \hat{\bf n}$ (where ${\bf{\hat{n}}} = {{\bf{{\nabla}}\eta} }/{|{\bf{{\nabla}}\eta}|}$ is the unit normal in the horizontal plane, see Fig. \ref{fig0}(b)) is included in the pressure calculation.  Moreover, $\boldsymbol{C}$ is the local phase-velocity of the wave, i.e., the horizontal speed of the surface's local vertical projection can be computed from $\eta(x,y,t)$ \citep{ayala2024}, as 
\begin{equation}
C_i = - \frac{\partial \eta}{\partial t} \, \frac{\partial \eta}{ \partial x_i } \,\frac{1}{|\boldsymbol \nabla \eta |^{2}} \, .
\label{eq:wavevel}
\end{equation} 
(Note that there are points for which  $C_i$  diverges, since  $|\nabla \eta|$ is in the denominator of \eqref{eq:wavevel}, but at any of these points $\alpha=0$ and $\partial \eta/\partial x=0$, so there are no singularities when evaluating $\tau^{\text{w-p}}$). For a simple monochromatic wave propagating in the $x$-direction with a surface elevation given by $\eta = a\cos{(kx - \omega t)}$, the resulting velocity using Equation \eqref{eq:wavevel} simplifies to the phase speed, $C_1 = c = \omega/k$. Nonetheless, Equation \eqref{eq:wavevel}  is formulated to remain valid for more general cases, including temporally evolving surfaces like broadband wave fields. 
Finally, the Heaviside function $\text{H}(x)=  \frac{1}{2}(x +|x|)/x$ is used to impose the pressure force only on the windward side of the wave. This approach follows the development in   \citet{ayala2024}, where it was assumed that displacement of the streamlines causes a pressure drop or prevents pressure recovery on the leeside of the wave, such that the pressure force there can be neglected.  
 Studies by \cite{buckley2020} and \cite{veron2007} have observed phenomena such as flow separation, incipient separation, or non-separated sheltering on the leeside under various wind-wave conditions, lending empirical support to this assumption. While the idea of displaced streamlines or leeward separation leading to negligible leeside pressure forces similarly aligns with earlier analytical approaches to modeling air-wave interactions (e.g., \cite{Belcher_Hunt_1993,jeffreys1}), 
it is a strong assumption that may not strictly apply to all wind-wave cases, especially fast swell waves or waves moving in the opposite direction of the wind. 

The approach in \citet{ayala2024} assumes a reference velocity $u_\Delta$ at a vertical distance equal to the horizontal LES grid spacing $\Delta x$. Here we instead assume that the reference velocity ${\bf u}_\Delta$ is a constant
representing the mean air velocity at a height $\Delta$, representing some sufficiently far, fixed distance above the wave field. Following \citet{meneveau2024},  $\Delta$ is set to a multiple of the roughness (wave) amplitude. 
Since the mean velocity reference ${\bf u}_\Delta$ is assumed constant, without loss of generality we may align the $x$ axis with this mean air velocity (i.e. ${\bf u}_\Delta  =  u_\Delta {\bf i}, \,\,{\rm and}\,\, v_\Delta = 0 $). Then, realizing that $\hat{n}_i$ and $\boldsymbol \nabla \eta $ are in the same direction, we can expand Equation \eqref{eq:mosd_wpm} to obtain the following kinematic wall stress in the $x$-direction:
\begin{equation} 
\tau^{\text{w-p}}_{xz}  =   u_\Delta^2\, \Biggl\{\frac{\alpha}{(\pi\,+\,\alpha)}\,\, \left(\left[\left( 1- \frac{C^+_x}{u^+_\Delta}\right) \, \hat n_x\right]^2 + \left[ \frac{C_y^+}{u_\Delta^+} \hat{n}_y \right]^2 \right)\,\, \frac{\partial \eta}{\partial x} \,\,  \text{H} \left[\left( 1- \frac{C^+_x}{u^+_\Delta}\right) \, \hat n_x -\frac{C_y^+}{u_\Delta^+} \hat{n}_y \right] \Biggr\},
\label{eq:press_stress}
\end{equation} 
where the subscript $(\cdot)^+$ denotes normalization with friction velocity $u_*$. For ease of exposition, in the remainder of the paper we use subscripts  $x$ (streamwise direction) and $z$ (vertical direction) instead of index notation. To model $\tau^{\text{w}-u}_{xz}$, i.e., tangential viscous stress contributions as well as unresolved surface form drag effects (e.g. from ripples), we use  
the friction factor $C_f$ parameterization: 
\begin{equation} 
\tau^{\text w-\nu}_{xz}  =  \frac{1}{2}\,\, C_f(Re_\Delta, z_0^{u}) \,\,u_\Delta^2.
\label{eq:visc_stress}
\end{equation}
The friction factor $C_f$ can be determined using the generalized Moody diagram fit developed by \citet{meneveau2020}, that depends on $Re_\Delta = u_\Delta \Delta /\nu$ and parameters representing atmospheric stability and unresolved roughness as follows:
\begin{equation}
C_f\,(Re_{\Delta},z_0^u) = 2\Biggl\{ \left[\frac{1}{2\,}C_{fs}(Re_\Delta) \right]^3 + \left[\frac{1}{\kappa} \left( \ln{\frac{\Delta}{z_0^u}} \,-\,\psi \right) \right]^{-6} \Biggr\}^{1/3}.
\label{eq:cftotal}
\end{equation}
Here $C_{fs}(Re_\Delta)$ is the smooth-surface friction coefficient that has been fitted to results from numerical integration of the equilibrium model differential equation \citep{meneveau2020}. The fit in it's simplest form (as also used in \citet{meneveau2024}) is given by $C_{fs}(Re_\Delta) = 0.0288\,Re_\Delta ^{-1/5}\,(1+577\,Re_\Delta^{-6/5})^{2/3}$. Also in Equation \eqref{eq:cftotal},   $\kappa = 0.4$ is the von Karman constant, and $Re_\Delta = u_\Delta \Delta /\nu = u_\Delta^+ \,\Delta^+$. 
Moreover, $\Delta^+ = \Delta u_*/\nu$, so for any given $u_*$ it (and $Re_\Delta$) can be computed from  the height $\Delta$ and fluid viscosity $\nu$. The second term in Equation \eqref{eq:cftotal} represents the effects of unresolved small-scale sea surface features and stability conditions of the airflow. The drag from unresolved surface features such as capillary waves that cannot be evaluated numerically on the  discretized surface grid is included by means of a small-scale roughness length $z_0^u$. This length can be expressed in terms of the root-mean-square surface fluctuations below the resolution with which $\eta(x,y,t)$ is known. Using the result from \citet{Geva} we define  $z_{0}^u = \eta^\prime_{\text{\rm sgs}} \, e^{-8.5 \kappa}$, where $\eta^\prime_{\text{\rm sgs}}$ is the root mean square of the sub-grid surface height distribution. The $z_{0}^u $-dependent term in Equation \eqref{eq:cftotal} vanishes ($z_{0}^u \to 0$) for surfaces that are known to be smooth below the resolved elevation 
field. We  also include effects due to airflow stability conditions through the stability function $\psi$ described in \citet{Barthelmie1999} as: 
\begin{equation}
    \psi = -5 \, \frac{\Delta}{L}, ~~~~ {\rm for} ~~~~ L>0,
    \label{eq:stable}
\end{equation}
and 
\begin{equation}
\psi = 2 \ln\left( \frac{1 + \chi}{2} \right) + \ln\left( \frac{1 + \chi^2}{2} \right) - 2 \tan^{-1} \chi + \frac{\pi}{2}, ~~~~{\rm with}~~~~~
    \chi = \left[ 1 - 16 \left( \frac{\Delta}{L} \right) \right]^{0.25},~~~{\rm for} ~~~~ L<0,
    \label{eq:unstable1}
\end{equation}
where $L$ is the Obukhov length.

The total horizontal drag over a surface of area $A$ is given by
\begin{equation} 
u_*^2 \, A = \langle \tau^{\text w-p}_{xz}+\tau^{\text w-\nu}_{xz} \rangle_{x,y} \, A = u_\Delta^2 \,\Lambda\,\,  A  \,\,\,\,\,\, \Rightarrow  \,\,\,\, u_\Delta^+ = \Lambda^{-1/2},
\end{equation}
where, similarly to the definition of the ``wind-shade factor'' of \citet{meneveau2024}, we  define the factor $\Lambda$ implicitly according to 
\begin{equation}
\Lambda = \Biggl <\frac{\alpha}{\pi\,+\,\alpha}\,\, \left(\left[ ( 1-  {C^+_x}\Lambda^{1/2}) \, \hat n_x\right]^2 + \left[  {C_y^+}\Lambda^{1/2} \, \hat{n}_y \right]^2 \right)\,\, \frac{\partial \eta}{\partial x} \,\,  \text{H} \left[(1-C_x^+\,  \Lambda^{1/2} ) \hat{n}_x - {C_y^+}\Lambda^{1/2} \, \hat{n}_y\right] \Biggr >_{x,y}+\, \frac{1}{2}\,C_f,
\label{eq:lambda}
\end{equation}
where $C_f=C_f(\Lambda^{-1/2} \Delta^+,z_{0}^u )$ and we have used $u_\Delta^+=\Lambda^{-1/2}$ so that $Re_\Delta = u^+_\Delta \Delta^+ = \Lambda^{-1/2} \Delta^+$. 
The slope $\partial \eta/\partial x$, angle $\alpha$, local surface normal vector ${\hat n}_i$, and local phase speeds $C_i$ are  all quantities that depend on position ($x,y$) and time. If the area $A$ is large enough, the planar averaging over $xy$, denoted by brackets $\langle \cdot \rangle_{x,y}$, is expected to converge to a well defined value of $\Lambda$ even for a single snapshot of a realization of the wave field (however, note that to determine ${\bf C}$ the vertical surface speed $\partial \eta/\partial t(x,y) $ is also required, and can be approximated by two consecutive snapshots of $\eta$ in time).
Solving for $\Lambda$ requires averaging over a known surface elevation map $\eta(x,y,t)$  numerically (except for some monochromatic waves, as discussed in Section 3\ref{sec:swarls} below). For a given value of $\Lambda$ (or $u_\Delta^+$) obtained during an iteration step from  Equation \eqref{eq:lambda}, the roughness length $z_0$ can be determined from $u_\Delta^+ = \kappa^{-1}\,(\ln{(\Delta/z_0)} - \psi)$, leading to:
\begin{equation}
z_0 = \Delta \exp{\left[-\kappa\,(\Lambda^{-1/2}+\psi)\right]} .
\label{eq:z_0}
\end{equation}
The reference height $\Delta$ is chosen following \citet{meneveau2024}, as $\Delta = 3H_p'$, where the typical dominant positive height of the surface \(H_p'\) is defined and computed as $H_p' = \langle[\rm max(0,\eta')]^8 \rangle^{1/8}$ where 
$\eta' = \eta - \langle \eta \rangle$ is the surface elevation relative to the mean elevation. This dominant height, $H_p'$, approximately represents the maximum positive deviation above the mean surface height and is closely related to typical characteristic wave heights, such as the significant wave height $H_s$. The choice of $H_p'$ ensures a consistent measure of the ``typical'' or representative maximum positive wave height across different wave conditions. 

If the friction Reynolds number $Re_\tau=u_* h/\nu$ is imposed (as is often the case in numerical simulations), the value of $u_*$ can be specified \textit{a priori} for a given boundary layer height $h$ and air viscosity $\nu$. Since  local phase speed $C_i$ can be determined locally at each point for a given surface, the dimensionless value $C^+_i=C_i/u_*$ can be evaluated at each point of the surface and used in the evaluation of $\Lambda$. For numerical convenience we clip the phase velocity $C_i$ 
when $|\nabla \eta|$ tends to zero in Equation \eqref{eq:wavevel} to $|{\bf C}|_{\rm max}=\sqrt{g/(0.25k_p)}$.
Using $0.25k_p$ ensures that the cutoff corresponds to speeds significantly faster than the fastest expected waves, of a size $4\times$ the wavelength of the peak wave.

In many cases, however, $u_*$ may not be known \textit{a priori}. In cases where e.g., the air velocity $U(h_r)$ at some reference height $h_r$ (e.g., the common choice $h_r=10$m) is known, an initial guess for $u_*$ can be used to evaluate $\Lambda$ via Equation \eqref{eq:lambda}. This value is then used to evaluate $u_* = \kappa U(h_r) / \log(h_r/z_0) = \kappa U(h_r) /[\log(h_r/\Delta) + \kappa\Lambda^{-1/2}]$. With a new value for $u_*$ thus determined, a next value of $\Lambda$ is computed, using Equation \eqref{eq:lambda}. Once converged (typically very fast), the final value of $z_0$ is determined using Equation \eqref{eq:z_0}.

As a cautionary remark, we stress that pressure stress model from \citet{ayala2024} (which is the basis of the SWARL model)  relies on the assumption that a horizontally discretized wavy surface can be effectively represented as a series of unconnected straight ramps, with the flow impinging horizontally and independently on each ramp. In reality, however, wind-wave dynamics near the crest exhibit flow accelerations and a drop in pressure to even negative values that can significantly affect drag \citep{Sullivan_2018}.  Such behavior cannot be accurately captured by the ``purely local'' potential flow model  invoked in the proposed methodology. A more realistic representation of near-surface flow, even within the framework of potential flow, would however require incorporating non-local effects, for instance, by solving partial differential equations using, e.g.,  eigenfunction expansions. Such methods are far more costly than the presently proposed  $z_0$ modeling framework requiring only a simple area integration over locally defined variables. Despite these simplifying assumptions, the results herein suggest that the proposed model—while offering a rather approximate description of wind–wave interactions—can still predict a global 
$z_0$ more accurately than existing models. 

\section{Description of datasets}

\begin{figure}[h]
 \centerline{\includegraphics[width=37pc]{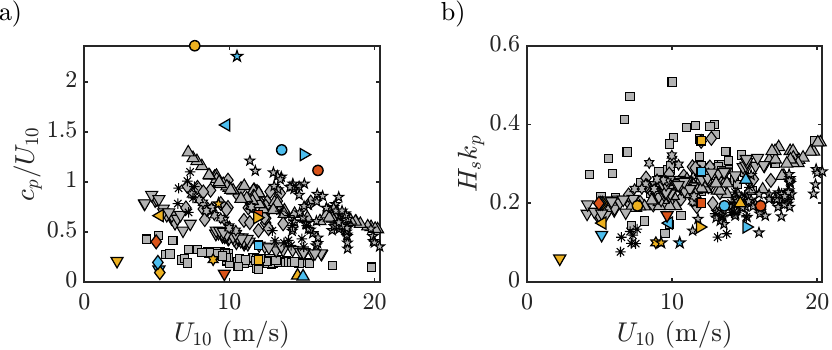}}
  \caption{Air-sea conditions for all 365 cases. a) Wave age vs velocity at 10-m height ($U_{10}$). b) Wave steepness vs $U_{10}$. Since the DNS/WRLES/Lab cases are monochromatic waves, they are plotted with $ak$, where $a$ is the amplitude of the wave. Symbols are as described in Table 1 and in Tables 1-11 from Supplementary Material}
  \label{fig1}
\end{figure}

This section describes the 365 separate data of unique air–sea interaction cases used to validate the SWARL model. These include 21 datasets comprised of Direct Numerical Simulations (DNS), Large Eddy Simulations (LES), and laboratory experiments, where the wavy surface elevation is fully characterized. The remaining datasets consist of field campaign measurements, which, while lacking full surface elevation profiles, report key wave parameters such as peak wave parameter and significant wave height.  Each case is treated as an independent realization of wind-wave conditions. The range of these conditions is illustrated in Figure~\ref{fig1}, and a detailed description of each dataset is provided below.

\subsection{DNS, LES and Laboratory experiments of wind over waves}

DNS and LES are high-fidelity simulation methods commonly used to study turbulent flows over complex surfaces such as waves. DNS resolves all turbulent scales without modeling, but its high computational cost restricts it to low-to-moderate Reynolds numbers. LES reduces computational demands by parameterizing the smaller turbulent subgrid-scales, resolving only the large-scale motions. In LES, the near-wall region can either be explicitly resolved (wall-resolved LES, or WRLES), or modeled (wall-modeled LES, or WMLES). WRLES applies a no-slip boundary condition at the wavy surface using a highly refined grid, while WMLES relies on parameterizations typically based on Monin–Obukhov Similarity Theory (MOST) \citep{moeng1984}, offering significant computational savings. Both WRLES and DNS simulations over wavy surfaces typically use boundary-fitted or terrain-following grids to capture the surface geometry \citep{Deskos2021}. For this study, we selected WRLES data from \citet{Wang_2021, Zhang2019, cao_shen_2021} and \citet{Hao2021}, where turbulent airflow is simulated over smooth monochromatic waves using time-dependent, surface-fitted grids. We also include DNS data from \citet{wu_popinet_deike_2022}, who employed a geometric Volume of Fluid method to capture the coupled wind–wave interface under various wind-wave conditions. In addition, we incorporate WMLES data of turbulent flow over 
 multiscale wave fields from \citet{Yang_Meneveau_Shen_2013} and \citet{sullivan2014}. Laboratory experiments are also included in this group. Specifically, we use the laboratory measurements of turbulent airflow over wind-generated waves from \citet{buckley2020} and \citet{yousefi_2020}. For the laboratory experimental datasets, the waves can be considered to be monochromatic waves with no capillary waves (ripples).  This group of DNS, WRLES and laboratory studies comprises 21 data sets. 

\begin{figure}[h]
 \centerline{\includegraphics[width=37pc]{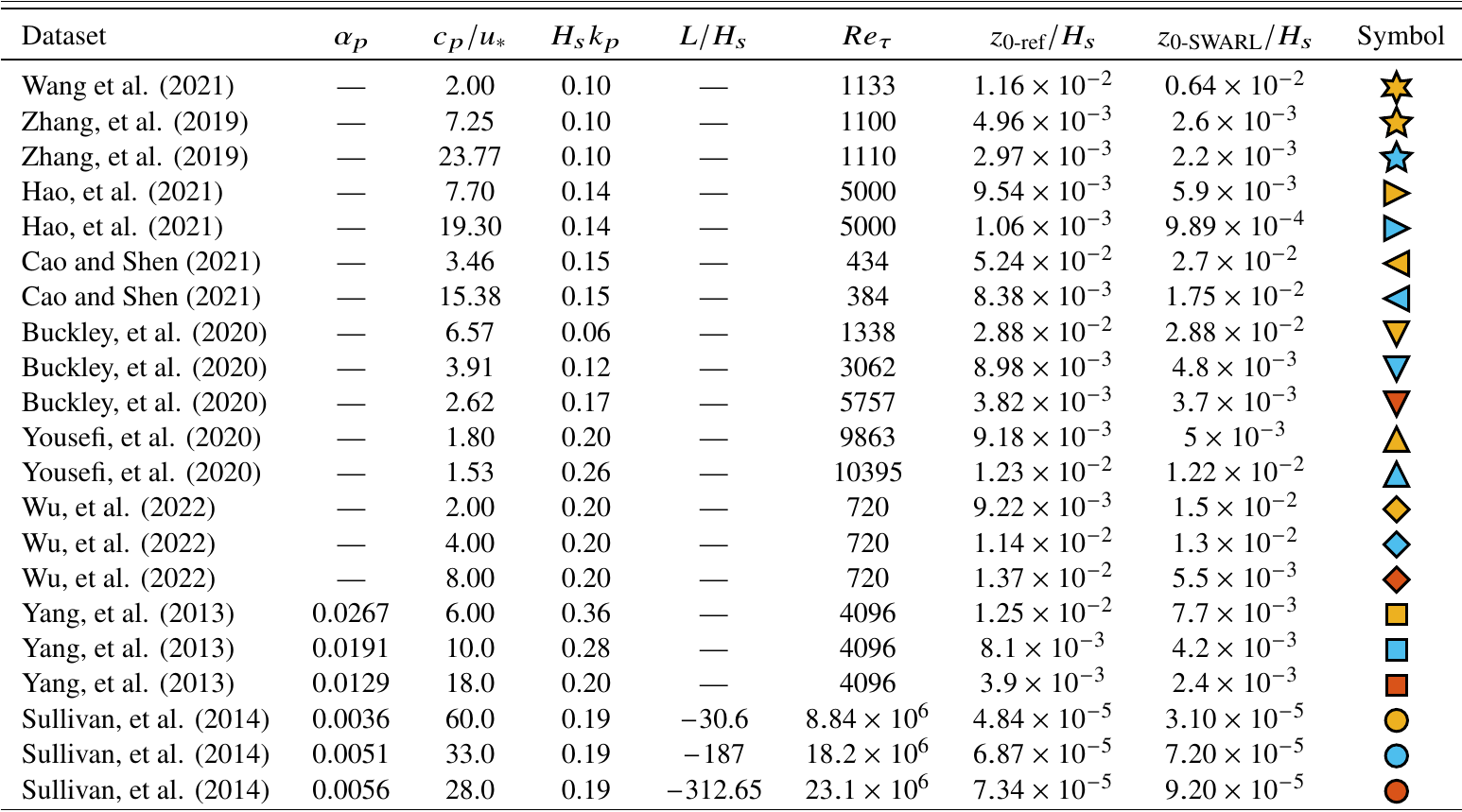}}
  \caption{DNS, WRLES and Laboratory experiments of wind over waves: summary of wind-wave parameters. For the monochromatic wave cases,  we take $H_s$ to be the amplitude of the wave ($a$), and the reference surface roughness is normalized by the wave amplitude. $z_{0\text{-ref}}$ represents the measured (ground truth) value of the roughness length, while $z_{0\text{-SWARL}}$ is the roughness length resulting from the SWARL model. For all cases, in this group the wind and wave directions are aligned.}
  \label{tab:wave_turbulence_data}
\end{figure}

In this group of DNS, LES and laboratory studies,  the exact surface distribution of the wavy surface is clearly defined by the authors (fully-characterized wave surface), as monochromatic waves \citep{Wang_2021, Zhang2019, cao_shen_2021, Hao2021,buckley2020,yousefi_2020}, third-order Stokes waves \citep{wu_popinet_deike_2022} or mutliscale waves \citep{Yang_Meneveau_Shen_2013,sullivan2014},  with known parameters. For all the cases  in this group, except the cases from \citet{sullivan2014}, neutrally stable conditions are reported. The \citet{sullivan2014} cases correspond to  weakly unstable conditions. The relevant wind-wave parameters of all the cases described in this group are compiled in Figure \ref{tab:wave_turbulence_data}. 

The reference surface roughness ($z_{0-\text{ref}}$) shown in Figure \ref{tab:wave_turbulence_data} was either reported by the authors or estimated by taking the reference velocity profiles and fitting the log law region with a least square method. The friction Reynolds number ($Re_\tau=u_*\delta/\nu$, where $\delta$ is the boundary layer height) is as reported from the authors, except for the \citet{sullivan2014} cases, which only reported $u_*$ and $\delta$ ($z_i$) in Table 1 of their article. For that case, the friction Reynolds number is estimated by utilizing their reported values and $\nu=1.5\times10^{-5}$ $\text{m}^2$/s for air. 

For all cases, the wave or peak wave (for multiscale wave) is moving parallel to and in the same direction as the airflow. The multiscale wave cases from \citet{Yang_Meneveau_Shen_2013} report the Phillips constant $\alpha_p$ (to be used to construct wave-field realizations based on the spectrum), while for the cases from \citet{sullivan2014}, the constant was estimated using the empirical relation:
\begin{equation}
    \alpha_p =0.006(U_{10}/c_p)^{0.55},
    \label{eq:alpha_p}
\end{equation}
from \citet{donelan1985}, where $U_{10}$ is the velocity at height of 10 m from the surface of the ocean wave. It is worth mentioning that some of the cases from \citet{sullivan2014} and \citet{cao_shen_2021} qualify as swell-dominated conditions based on their wave age $c_p/U_{10} > 1.2$ (see Fig.~\ref{fig1}). However, we have found that from their respective studies there is no evidence of wave-to-wind momentum transfer (i.e., thrust) when comparing the mean velocity profiles to those over a flat, smooth wall at the same Reynolds number. Therefore these data are included in our comparison as well.

\subsection{Field campaign experiments of wind over waves}

The remaining datasets come from field experiments that provide real-world observations of turbulent airflow over ocean waves. Our data selection prioritized studies offering comprehensive wind-wave parameters, specifically peak wave characteristics ($k_p$, peak angular frequency $\omega_p$, or peak frequency $f_p$), significant wave height $H_s$, friction velocity $u_*$, wind speed at 10 m height $U_{10}$, atmospheric stability as indicated by the Obukhov length scale $L$, and the orientation between wind and waves.

The Grand Banks ERS-1 SAR Wave Spectra Validation Experiment \citep{GrandBanks} provided detailed measurements of wind stress and directional wave spectra in open ocean conditions, based on measurements from a bow anemometer system aboard a research vessel. All data points from Table 1 of \citet{GrandBanks} were included for model validation and compiled in Table S1 of the Supplementary Material. Neutral stability conditions were assumed by the original authors.

The Humidity Exchange over the Sea Main Experiment (HEXMAX) \citep{hexmax} utilized sonic and pressure anemometers for simultaneous measurements of wind and wave parameters, focusing on the drag coefficient and roughness length relationships. Only HEXMAX cases with complete parameter sets from sonic anemometer measurements were selected from \citep{hexmax} and are compiled in Table S2-S3 of the Supplementary Material, with neutral stability conditions assumed following the authors' approach.

The Lake George study by \citet{lakegeorge} investigated how wind trends and sub-minute gustiness affect the air-water drag coefficient over a fetch-limited lake. Their analysis employed tower-based sonic-anemometer turbulence profiles complemented by co-located pressure plates, wave staffs, and directional wave spectrometers. We selected all cases from Table 1 of \citet{lakegeorge}, which are summarized in Table S4-S5 of the Supplementary Material. Although the authors focus was on wind gustiness effects, in our analysis we exclude gustiness and wind trend parameters. The authors' assumptions of neutral stability and aligned wind-wave conditions is maintained.

The Lake Ontario experiment by \citet{lakeontario} focused on air-water momentum flux over shoaling waves using multiple towers at different depths, highlighting the effects of wave steepness and celerity. All reported cases from \citet{lakeontario} were selected and the relevant details are compiled in Table S6 of the Supplementary Material, assuming neutral stability conditions based on small Richardson numbers and aligned wind-wave conditions as per the original study.

The Risø Air-Sea Exchange (RASEX) experiment \citep{rasex} utilized sonic anemometers at various heights and wave instruments in shallow coastal waters of Denmark, examining sea-surface roughness dependency on wind-generated waves. All wind-wave cases from Table 1 of \citet{rasex} were chosen, with relevant parameters summarized in Table S7-S8 of the Supplementary Material, maintaining the original assumption of wind-wave alignment.

The Gulf of Tehuantepec Experiment (GOTEX) \citep{gotex} involved airborne measurements of fetch-limited waves during strong offshore winds, including detailed surface wavenumber spectra and turbulent fluxes. Due to the inherent uncertainties involved in matching wind parameters with spatial spectra, we limited our selection to summary data from Tables 1 and 2 of \citet{gotex}. These parameters are presented in Table S9-S10 of the Supplementary Material.

The Surface Wave Dynamics Experiment (SWADE) \citep{swade, swade2} provided direct measurements of momentum, heat, and water vapor fluxes alongside directional wave spectra off the coast of Virginia. The study aimed to understand the modification of drag coefficients due to counter- and cross-swell interactions. Importantly, from \citet{swade} we specifically extracted peak wave frequency, and from \citet{swade2} we obtained $u_*$, $U_{10}$, and $H_s$. Only cases representing pure wind-sea conditions aligned with wind direction were selected. These cases are summarized in Table S11 of the Supplementary Material.

Across all datasets, at least one peak wave parameter ($k_p$, $\omega_p$, $f_p$) was available. Where necessary, the deep-water dispersion relation ($\omega = \sqrt{gk}$) was used to compute additional wave parameters. Surface roughness lengths from the RASEX, Lake Ontario, and Grand Banks studies were adopted directly. For the GOTEX, Lake George, SWADE, and HEXMAX datasets, surface roughness length ($z_{0-\text{ref}}$) was estimated using:
\begin{equation}
z_{0-\text{ref}} = h_{10} \exp\left(-\frac{\kappa}{\sqrt{C_D}}\right),
\end{equation}
where $h_{10} = 10$ m, and the drag coefficient $C_D$ is defined as $(u_*/U_{10})^2$. The Phillips constant, $\alpha_p$ for each dataset was computed using Equation \eqref{eq:alpha_p}. Since only the $u_*$ is reported in  the field campaign experiments, we estimate the friction Reynolds number for all of these cases using $Re_\tau=u_*\delta/\nu$ with $\delta=1000$m and $\nu=1.5\times10^{-5}$ $\text{m}^2$/s. 

\subsection{Generation of time-dependent wavefields to enable application of the SWARL model}

As described in \S \ref{sec:wave_drag_model},  the SWARL model requires as input, realizations of the surface height distribution $\eta(x,y,t)$. For the cases of monochromatic wave, the wave height distribution can be simply generated by evaluating a single-mode harmonic wave according to $\eta(x,y,t) = a \cos{(kx -\omega t)}$. 

For  multiscale wavefields, the wave height distribution $\eta(x,y,t)$ is generated by the traditional method of superposing random-phase traveling waves with a prescribed surface spectrum $S(k_x,k_y)$. We generate the multiscale surface utilizing a standard assumed wave spectrum model. Specifically in this work we use the JONSWAP spectrum \citep{hasselman1973}:
\begin{equation}
S(k_x,k_y) = \frac{\alpha_p}{2 k^4} \exp{\left[ -\frac{5}{4} \left( \frac{k_p}{k}\right )^2 \right]}\,\, 3.3^ \gamma\,\,D(\theta),
 \label{eq:jonswap}
\end{equation}
where $\gamma={\exp{[-{1}/{(2 \varepsilon^2)} ( \sqrt{{k}/{k_p}}-1)^2  }}]$ and directionality is prescribed by adopting the widely used spreading function \citep{hasselmann1980,cartwright1963}
$D(\theta) = ({2}/{\pi}) \cos^2{(\theta - \theta_w)}$ if $|\theta | \leq {\pi}/{2}$, where $\theta = \arctan{(k_y/k_x)}$ and $\theta_w$ is the direction of the peak wave. 
The numerically generated surfaces used a $N_x=1280 \times N_y=1280$ grid with domain size $L_x = L_y = 10\lambda_p$
i.e., $\Delta x = \Delta y = 10\lambda_p/N_x = 0.007812 \, \lambda_p$. Two successive time instants were stored, separated by a time   interval $\Delta t$ sufficiently small to enable accurate evaluation of time-derivatives of the surface elevation using first-order finite differencing (the dimensionless time interval differed between cases, with the smallest being $\Delta t \,c_p/\Delta x = 0.008$ and the largest 
$\Delta t \,c_p/\Delta x = 0.13$). A grid-independence study using the current methodology found no significant changes in the results using finer grid or temporal resolutions. Spatial and temporal gradients of all surfaces were computed using  first-order finite differencing.  We note that this approach to reconstruct multiscale wave fields does not reproduce the exact statistical features of wave surfaces of the  selected datasets since the full 2D wave spectra are not typically available. Nonetheless, we verified that the synthesized surfaces reproduce key wave statistics, particularly the significant wave height $H_s$, with reasonable fidelity. 

\begin{figure}[h]
 \centerline{\includegraphics[width=37pc]{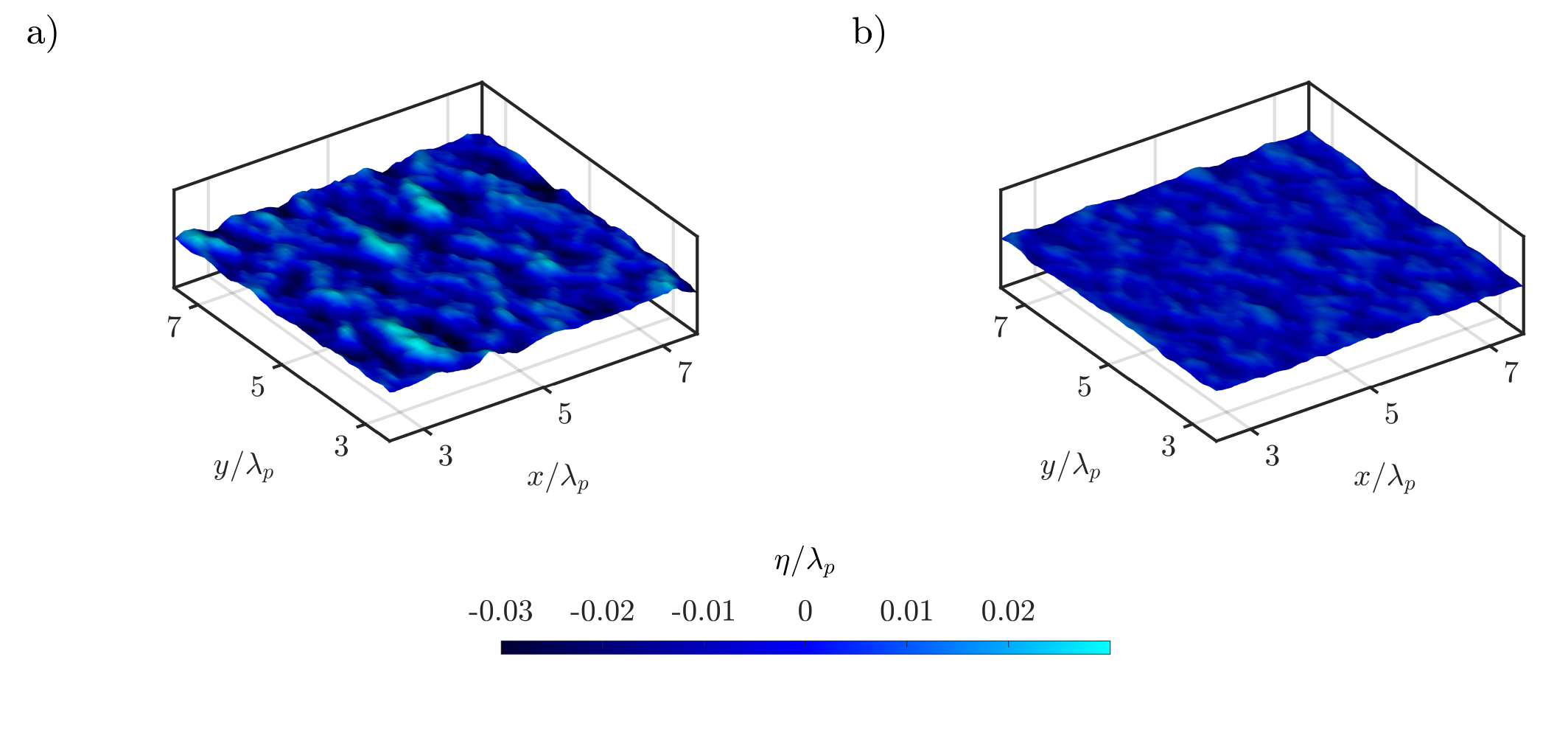}}
  \caption{(a) Sample snapshot of a synthetically generated multiscale wavefield with $c_p/u_*=10$  from \citet{Yang_Meneveau_Shen_2013}, and (b) for the dataset from  \citet{rasex} for $c_p/u_*=19.1$. }
  \label{figsynthetic}
\end{figure}

Figure \ref{figsynthetic} displays two snapshots of example synthetically generated fields used to create the surfaces for the $c_p/u_*=10$ data from \citet{Yang_Meneveau_Shen_2013} (shown in panel (a)) and for the  $c_p/u_*=19.1$ data from \citet{rasex} (shown in panel (b)). 

Since using a numerical approach to generate the surface will make the surface inherently filtered at the grid resolution $(\Delta x, \Delta y)$, we include drag effects of the sub-filter features using a ``subgrid-scale'' roughness length $z_0^u$. For the JONSWAP model, in the equilibrium region where the spectral behavior follows the power-law $k^{-4}$, we note that  $\gamma \rightarrow0$ approximately at $k \approx 2\,k_p$, which is  typically much smaller than the resolution/filter wavenumber $k_{\Delta } = ((\pi/\Delta x)^2 + (\pi/\Delta y)^2)^{1/2}$. The r.m.s of the sub-filtered surface height fluctuations ($\eta_{\rm sgs}'$) is evaluated by integrating  the one-dimensional form of Equation \eqref{eq:jonswap} from $k_{\Delta}=\pi$ to $k_{\Delta}=\infty$, resulting in:
 \begin{equation}
\eta^{\prime}_{\text{sgs}} \,\,=\,\,\frac{\sqrt{0.2 \alpha_p}}{k_p} \left( 1 - \exp{\left[-\frac{5}{4}\left(\frac{k_p}{k_\Delta}\right)^2 \right] }\right)^{1/2}.
 \label{eta_rms}
\end{equation}
Once the r.m.s. of the subgrid height fluctuations is known, the roughness length can be evaluated according to \cite{Geva}:
$ z_0^u = \eta^{\prime}_{\text{sgs}}\,e^{-\kappa \,8.5}$
and used in Equation \eqref{eq:cftotal} to evaluate the friction factor to model the unresolved roughness as part of the SWARL model. We note in passing that only for the \citet{sullivan2014} cases, we set $ z_0^u= 0.0002$m, since this is the reported surface roughness value used by the authors to represent unresolved wave surfaces. 
 
\section{Results}

Once the surfaces were generated for all 365 unique wind-wave scenarios we apply the proposed SWARL model. For the iterative procedure to determine $\Lambda$ and $z_0$, we use a Newton-Rahpson method to obtain a converged solution of Equation \eqref{eq:lambda} using a tolerance of $10^{-6}$. Since our central claim is that the model can outperform current state-of-the-art approaches when provided with full knowledge of the wavy surface height distribution, we begin by evaluating its performance on cases with fully characterized surfaces. We then extend the comparison to include all available cases.

\subsection{Comparison with empirically-fitted models and algorithms}

The performance of SWARL is compared against several of the empirically-fitted models and algorithms described  in section \ref{sec:intro}. For the Charnock model \eqref{charnock} we use $\alpha_{\text{ch}} = 0.023$, which reflects the center of the reported range and for the WAM model we use $\hat{\alpha} =0.0185$. In this study, the COARE1  and COARE2 models estimate surface roughness using Equation~\eqref{coare1} and Equation~\eqref{coare2}, which correspond to the wave-age-based and wave-steepness-based parameterizations of $\alpha_{\text{ch}}$, respectively. The surface roughness models for ocean waves discussed in Section~\ref{sec:intro} account only for the pressure-drag (fully rough) component of the total roughness. To enable an appropriate comparison with the proposed model—which captures both rough and viscous contributions—a viscous ``smooth'' component should be added to all the comparison models. The total surface roughness length is therefore given by:
\begin{equation}
z_0 = z_0^{r} + 0.11\frac{\nu}{u_*},
\end{equation}
where, $z_0^{r}$ is the roughness predicted by the empirical models described before, and the second term represents the viscous contribution.  

\subsubsection{Fully-characterized surfaces} 

\begin{figure}[h!]
 \centerline{\includegraphics[width=37pc]{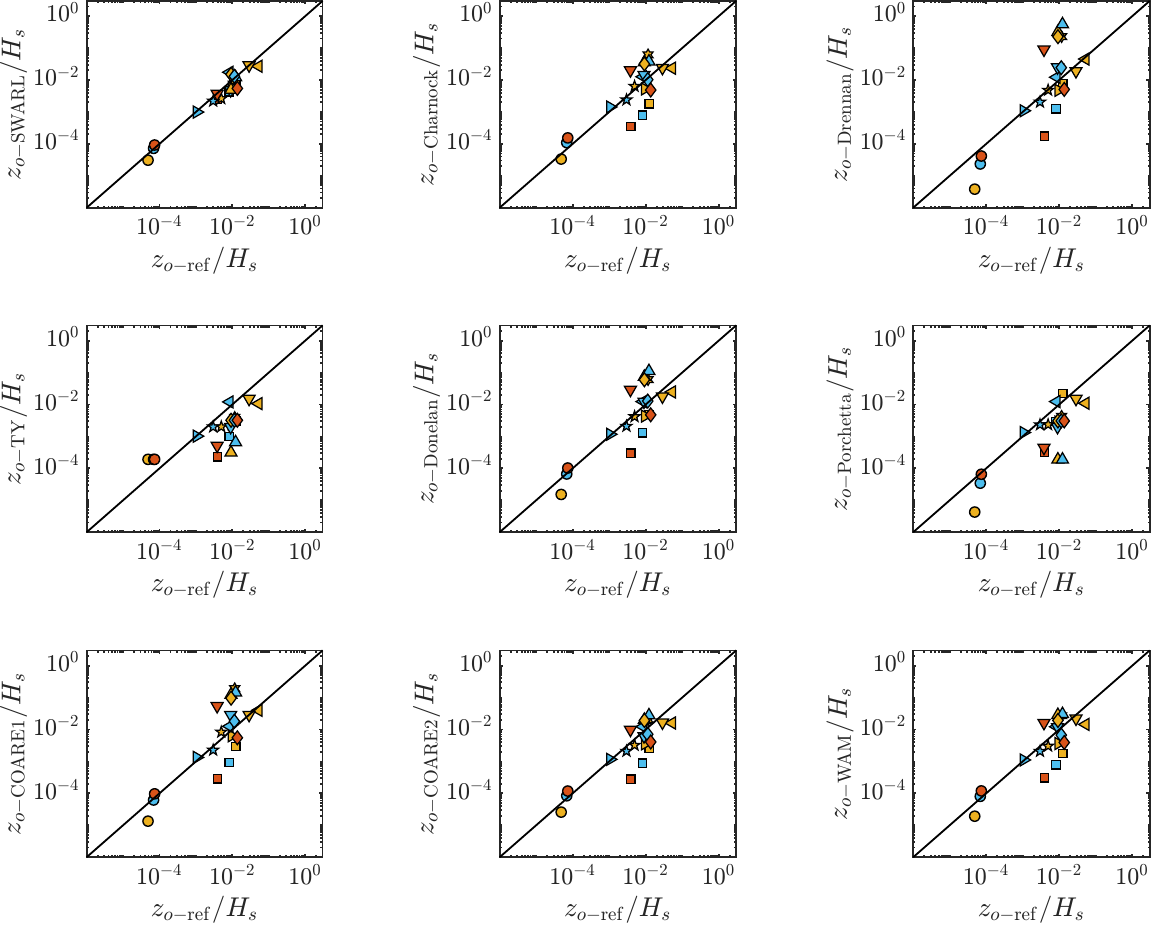}}

  \caption{Surface roughness length predicted by  models versus values from reference data for cases where the surface is fully-characterized. For monochromatic wave cases, the surface roughness is normalized by the amplitude $a$. Symbols are as described in Table 1}
  \label{fig2}
\end{figure}

Figure~\ref{fig2} presents scatter plots comparing modeled surface roughness lengths to reference data for the wind-wave cases where the wave is fully-characterized (see Figure \ref{tab:wave_turbulence_data} for details of these cases). Here, the SWARL model shows strong  agreement with the data and clearly outperforms the other models. To quantitatively assess performance, we compute the correlation coefficient, mean logarithmic error ($e_1 = \langle |\log_{10} \left({z_{0-\text{mod}}}/{z_{0-\text{ref}}}\right)|\rangle$), and mean absolute error ($e_2 = \langle |\left({z_{0-\text{mod}}/z_{0-\text{ref}}}\,-1\right)|\rangle$) for each model (see Table~\ref{t1}). The SWARL model achieves nearly twice the correlation of all other models except the Taylor-Yelland model. However, the Taylor-Yelland model exhibits a logarithmic error nearly three times higher  and nearly double the absolute error compared to SWARL. Note that the significant $e_2$ error exhibited by the Porchetta model is entirely due to the values of A and B  obtained by the authors from fitting to a data sample for which angular dependence was available, i.e., different data sets from the ones presented here. Moreover, the larger B exponent (compared to the Drennan model) makes the model very sensitive to $c_p^+$. Improved correlations and smaller errors can be expected if these coefficients were to be fitted again based on the larger number of data samples used in the present study.

\begin{figure}[h]
 \centerline{\includegraphics[width=37pc]{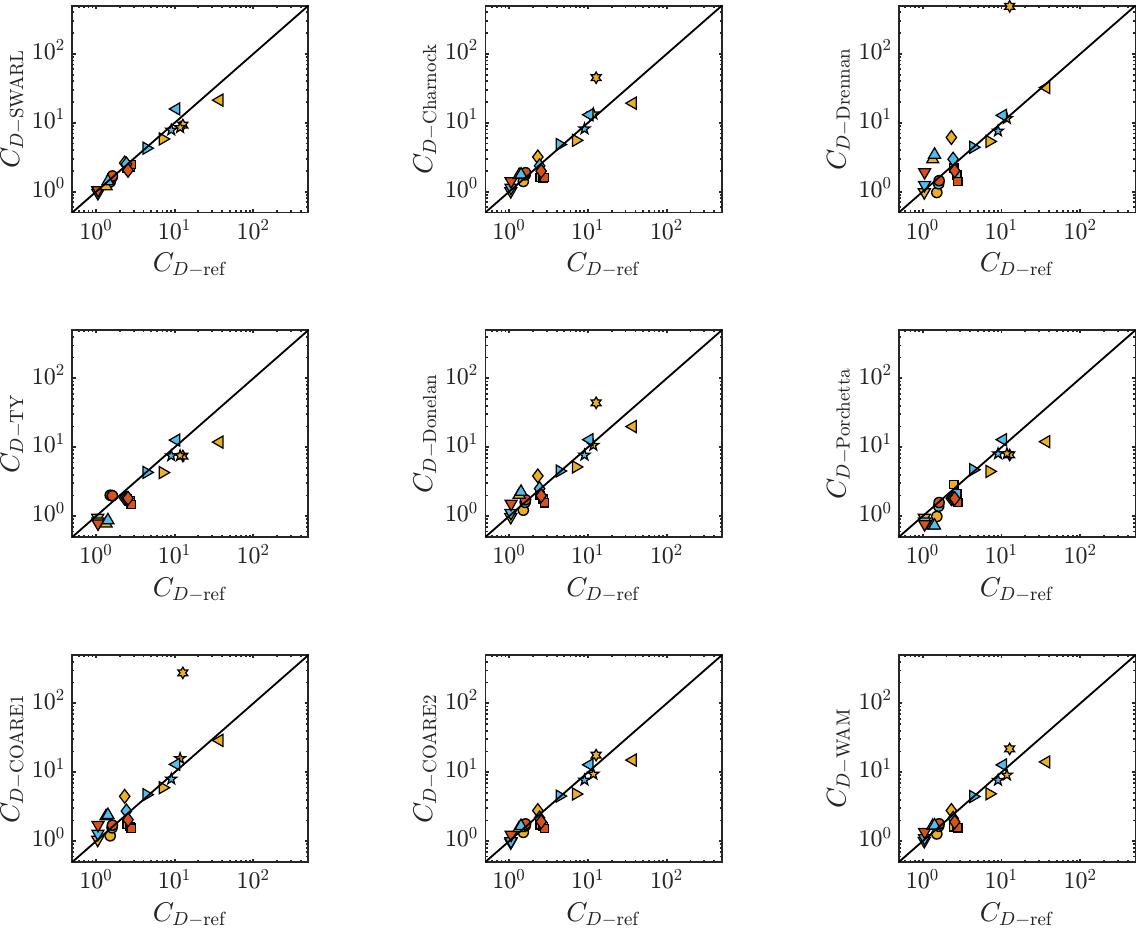}}
  \caption{Drag coefficient ($C_D \times 1000$) predicted by the models versus values from reference data for cases where the surface is fully-characterized. Symbols are as described in Table 1.}
  \label{fig3}
\end{figure}

An additional comparison can be made via the implied drag coefficient, which provides a more direct relation to drag forces:

\begin{equation}
C_D = \left( \frac{u_*}{U_{10}}\right)^2 = \left[ \frac{\kappa}{\ln{(10/z_0)} - \psi } \right]^2.
\end{equation}
We compute $C_D$ using the surface roughness values predicted by each model and compare them to the reference data in the scatter plots of Fig.~\ref{fig3}. Again, the SWARL model shows good agreement with the reference values. A quantitative comparison is provided in Table~\ref{t2}, where the SWARL approach consistently achieves the highest correlation amoung the models tested. The next-best performance comes from the Taylor-Yelland and COARE2 models, similar to the surface roughness results; however, the SWARL model results show about half of the logarithmic error compared to both of these approaches. Overall, the SWARL model reduces error by a factor ranging from 1.67 to 4 relative to the other models for the cases in Figure \ref{tab:wave_turbulence_data}.

\begin{table}[h!]
\caption{Model performance parameters for the modeled  surface roughness length $z_0$, for fully-characterized wave surfaces}\label{t1}
\begin{center}
\begin{tabular}{ccccc}
\topline
Model & $\rho$ & $e_1$ & $e_2\,\, $  \\
\midline
 SWARL & 0.839 & 0.178 & 0.352 \\
 Charnock & 0.348 & 0.408 & 1.168 \\
 Drennan & 0.080 & 0.646 & 7.283 \\
 Taylor-Yelland & 0.707 & 0.613 & 0.889 \\
 Donelan & 0.192 & 0.439 & 1.827 \\
 Porchetta & 0.055 & 0.804 & 33.540\\

 COARE1 & 0.171 & 0.525 & 3.343 \\
 COARE2 & 0.469 & 0.365 & 0.669 \\
 WAM & 0.381 & 0.399 & 0.833 \\
\botline
\end{tabular}
\end{center}
\end{table}

\begin{table}[h!]
\caption{Model performance parameters for the modeled drag coefficients, for fully-characterized wave surfaces}\label{t2}
\begin{center}
\begin{tabular}{ccccc}
\topline
Model & $\rho$ & $e_1$ & $e_2 \,\,$ \\
\midline
 SWARL & 0.926 & 0.064 & 0.137 \\
 Charnock & 0.611 & 0.125 & 0.333 \\
 Drennan & 0.263 & 0.215 & 2.174 \\
 Taylor-Yelland & 0.819 & 0.159 & 0.299 \\
 Donelan & 0.620 & 0.131 & 0.359 \\
 Porchetta & 0.704 & 0.262 & 1.490 \\

 COARE1 & 0.305 & 0.180 & 1.280 \\
 COARE2 & 0.807 & 0.107 & 0.216 \\
 WAM & 0.734 & 0.118 & 0.245 \\

\botline
\end{tabular}
\end{center}
\end{table}

The SWARL approach is based on modeling some aspects of the local physics of airflow over waves, and for this purpose it requires instantiation of surface height elevation (and evaluation of the local phase speed ${\bf C}$, which requires time derivatives). When the wave surface from the available datasets is well characterized,  the qualitative and quantitative comparisons in this section demonstrate that the SWARL model outperforms widely used empirical and state-of-the-art approaches for predicting surface roughness and drag coefficient over ocean waves.

\subsubsection{All surfaces} 
\begin{figure}[h]
 \centerline{\includegraphics[width=37pc]{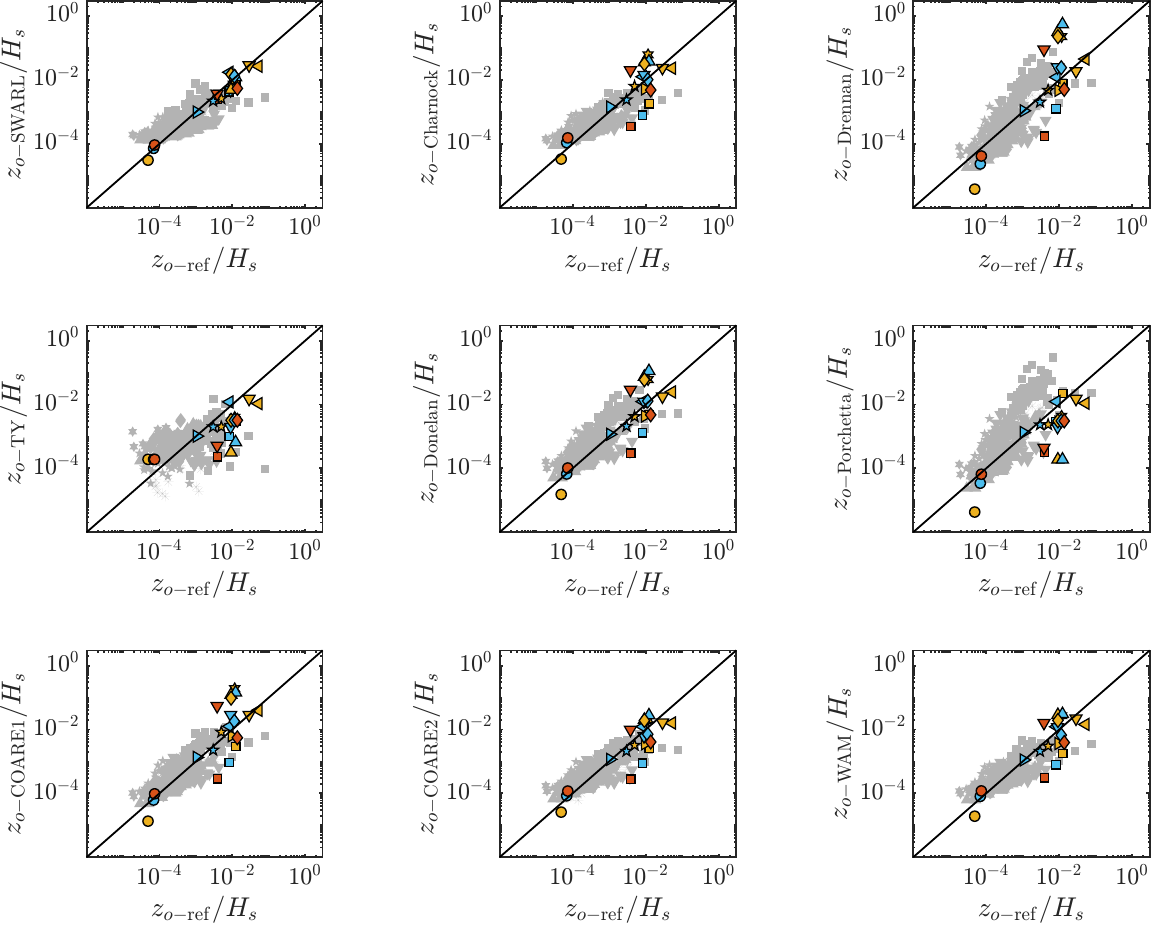}}
  \caption{Surface roughness length $z_0$ predicted by the model versus values from reference data for all surface cases. For monochromatic wave cases, the surface roughness is normalized by the amplitude $a$. Symbols are as described in Table 1 and in Tables S1-S11 from the Supplementary Material}
  \label{fig4}
\end{figure}

\begin{figure}[h]
 \centerline{\includegraphics[width=37pc]{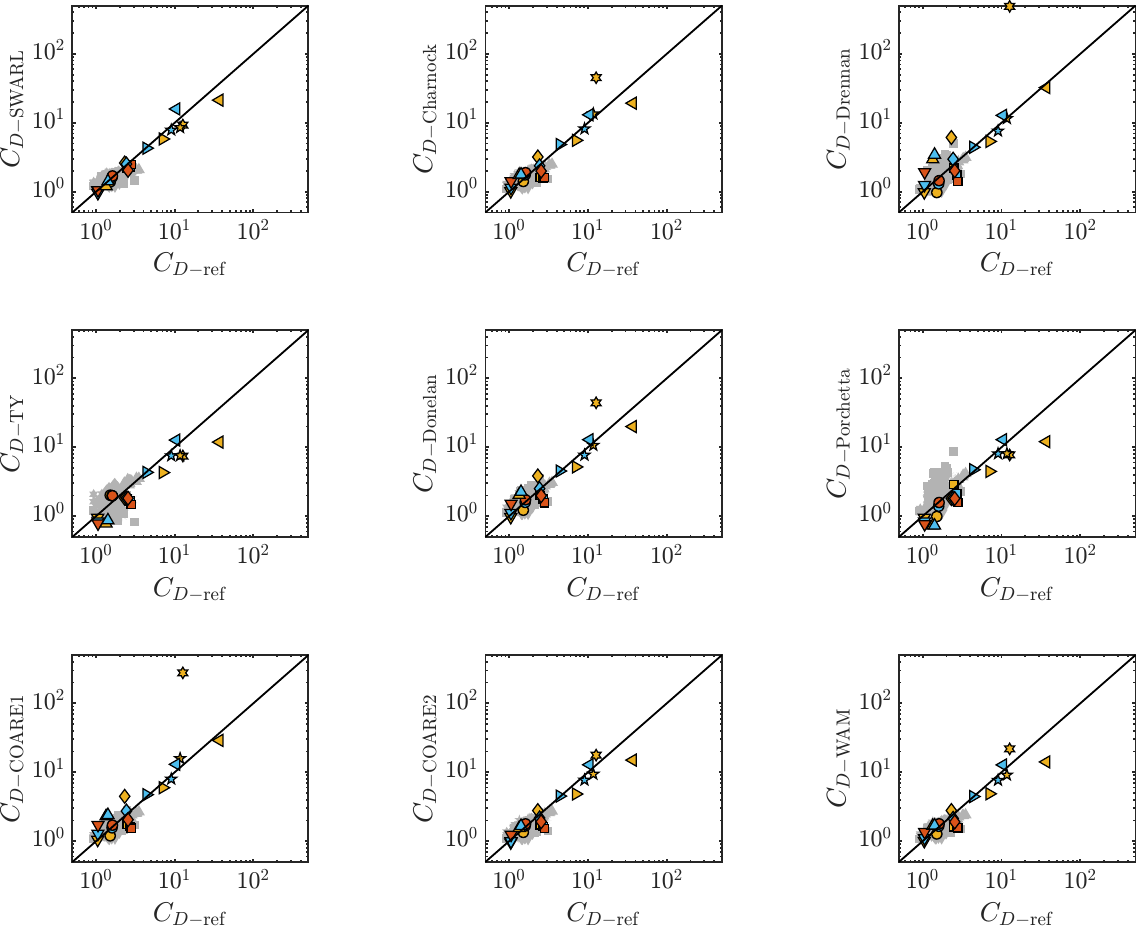}}
  \caption{Drag coefficient ($C_D \times 1000$) predicted by the model versus values from reference data for all surface cases.Symbols are as described in Table 1 and in Tables S1-S11 from Supplementary Material}
  \label{fig5}
\end{figure}

We now evaluate the performance of the models across all available datasets, including field data for which full quantitative wave-field information was not available. Figure~\ref{fig4} presents scatter plots comparing modeled surface roughness with reference data for a wide range of wind-wave conditions. Qualitatively, the SWARL, Charnock, COARE2, and WAM models show the best agreement with the reference values. Quantitatively, as shown in Table~\ref{t3}, the SWARL model achieves a logarithmic error and mean absolute error comparable to those of the Charnock, COARE2, and WAM models, but with a higher correlation coefficient. This trend is similarly observed in the drag coefficient predictions (Figure~\ref{fig5}) and the corresponding performance metrics in Table~\ref{t4}. While the Charnock and WAM models demonstrate good performance, it is important to note that they rely on the specification of empirical parameters—namely, the Charnock constant and the $\hat{\alpha}$ parameter, respectively. In contrast, the SWARL model requires no such tuning and instead uses realizations of the wave height distribution  as input to the numerical evaluation of the parameter $\Lambda$. We conclude that even when the wave surface is only approximately known, the SWARL model remains competitive or slightly superior to state-of-the-art models.

\begin{table}[h]
\caption{Model performance of surface roughness for all wave surfaces}\label{t3}
\begin{center}
\begin{tabular}{ccccc}
\topline
Model & $\rho$ & $e_1$ & $e_2\,\,$  \\
\midline
 SWARL & 0.593 & 0.316 & 1.032 \\
 Charnock & 0.411 & 0.308 & 1.003 \\
 Drennan & 0.228 & 0.432 & 2.225 \\
 Taylor-Yelland & 0.380 & 0.437 & 1.666 \\
 Donelan & 0.328 & 0.378 & 1.593 \\
 Porchetta & 0.199 & 0.613 & 7.538 \\

 COARE1 & 0.299 & 0.323 & 1.163 \\
 COARE2 & 0.470 & 0.298 & 0.878 \\
 WAM & 0.444 & 0.290 & 0.826 \\

\botline
\end{tabular}
\end{center}
\end{table}

\begin{table}[h]
\caption{Model performance of drag coefficient for all wave surfaces}\label{t4}
\begin{center}
\begin{tabular}{ccccc}
\topline
Model & $\rho$ & $e_1$ & $e_2\,\,$  \\
\midline
 SWARL & 0.931 & 0.065 & 0.151 \\
 Charnock & 0.680 & 0.064 & 0.154 \\
 Drennan & 0.333 & 0.094 & 0.331 \\
 Taylor-Yelland & 0.802 & 0.088 & 0.202 \\
 Donelan & 0.688 & 0.079 & 0.201 \\
 Porchetta & 0.737 & 0.140 & 0.458 \\

 COARE1 & 0.377 & 0.069 & 0.215 \\
 COARE2 & 0.842 & 0.061 & 0.141 \\
 WAM & 0.787 & 0.060 & 0.137 \\

\botline
\end{tabular}
\end{center}
\end{table}

\begin{figure}[h]
 \centerline{\includegraphics[width=37pc]{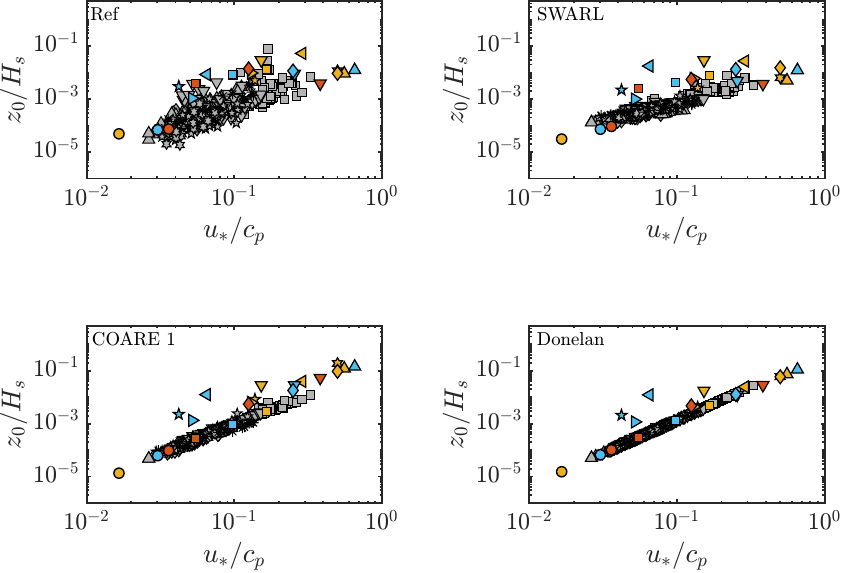}}
  \caption{Surface roughness vs. inverse wave age for all surface wave cases. Top left: Reference data, Top right: SWARL model, Bottom left: COARE1 model and Bottom right: Donelan model. Symbols are as described in Table 1 for the fully characterized wave-field cases (color symbols), while gray symbols represent the field data where surfaces are not fully characterized (more details provided in Tables S1-S11 from Supplementary Material).}
  \label{fig6}
\end{figure}

\begin{figure}[h]
 \centerline{\includegraphics[width=37pc]{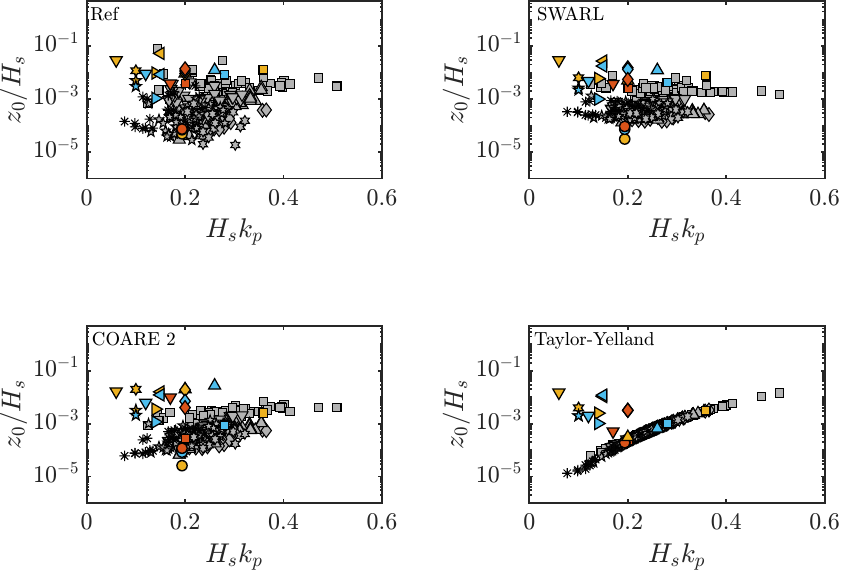}}
  \caption{Surface roughness vs wave steepness for all surface wave cases. Top left: Reference data, Top right: SWARL model, Bottom left: COARE2 model and Bottom right: Taylor-Yelland model. Symbols are as described in Fig. \ref{fig6}.}
  \label{fig7}
\end{figure}

To further illustrate the behavior of the proposed model across key wave-based scaling parameters, Figures~\ref{fig6} and \ref{fig7} present the estimated surface roughness length $z_0$ as a function of inverse wave age and wave steepness, respectively. In Fig.~\ref{fig6}, we include the COARE1 and Donelan models, both of which explicitly incorporate inverse wave age scaling in their formulations (Equations~\eqref{zo_waveage} and \eqref{coare1}). As expected, these models exhibit a clear power-law trend (linear in log-log plot). The SWARL model also displays an approximately linear trend, but with greater scatter, more closely resembling the variability seen in the reference data. This ability to capture the more realistic data spread is because the SWARL representation depends on a far greater number of surface characteristics (although the variability in the SWARL model results is still smaller than that of the data).

Figure~\ref{fig7} shows a similar comparison against wave steepness, where we include the COARE2 and Taylor-Yelland models—both of which rely on wave steepness scaling in their formulations, respectively in Equations~\eqref{coare2} and \eqref{TY}. The Taylor-Yelland model demonstrates a well-defined steepness scaling, while the COARE2 model does not. This is because, although the COARE2 model uses steepness to estimate the Charnock parameter, its full formulation (via Equation~\eqref{charnock}) leads to:
\begin{equation}
z_0/H_s = 0.09\left(u_*/c_p\right)^2,
\end{equation}
which effectively removes explicit dependence on steepness. In contrast, the SWARL model does not enforce a specific scaling and instead exhibits behavior that more closely matches the variability observed in the reference data, particularly under diverse sea states.

\subsection{Modified Charnock coefficient}

\begin{figure}[h!]
 \centerline{\includegraphics[width=37pc]{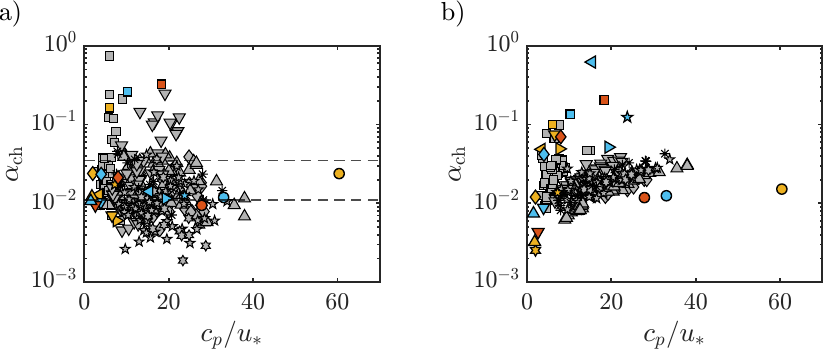}}
  \caption{Scatter plot of modified Charnock coefficient versus normalized wave velocity for  measured roughness lengths (a), and using Equation \eqref{eq:mod_charnock} (b). The lines represent the range of typical used Charnock constant values (0.011-0.038). Symbols are as described in Fig. \ref{fig6}.}
  \label{fig_charnock_const}
\end{figure}

In order to cast the comparison in terms of another common parameter, we also present results in terms of the effective (modified) Charnock coefficient:
\begin{equation}
\alpha_c = z_0 \, k_p (c_p^+)^2 = (\Delta\,k_p)\,(c_p^+)^2\,\exp{\left[-\kappa\,{\Lambda^{-1/2}}\right]},
\label{eq:mod_charnock}
\end{equation}
where the last equality is for the proposed model when $z_0$ is expressed in terms of $\Lambda$ and $\Delta$. Although numerous formulations exist for $\alpha_c$ , each parameterized by wave properties, no universal formulation has been established. Most are derived from data fitting \citep{zhao2024,linsheng2020}.  The modified Charnock coefficient was calculated for all wave cases and the results from data (using the measured values of $z_0$) are plotted in Fig.\ref{fig_charnock_const} alongside results from the SWARL model.

As can be seen, the measured results do not fall on a single curve depending only on wave age and instead show significant scatter since the parameter still depends on many additional wave surface characteristics. Similarly,    the modified Charnock coefficient derived from the proposed  model shows the desired lack of one-to-one dependence or single scaling with wave age. Rather, the model incorporates dependency on additional parameters beyond wave age, namely wave steepness (capturing surface geometric effects) and wind conditions (via $Re_\tau$).

\subsection{Simplified model (SWARL-S)}
\label{sec:swarls}

For monochromatic waves of the form $\eta = a\,\cos(kx-\omega t)$ with small $(ak)$ values, the calculation of   $\Lambda$   can be simplified. In particular, for $\alpha \ll\pi$ and $\alpha \approx \partial \eta/\partial x$, $\alpha/(\pi + \alpha) \approx  \pi^{-1} \partial \eta/\partial x$, $C=\omega_p/k_p=c_p$   is constant everywhere, and therefore the  wave-history term $1-C^+\sqrt{\Lambda}$ is also constant everywhere. Noting that the average of $(\partial \eta/\partial x)^2$ over a half wavelength (since we only consider forward-facing side generates drag as discussed in Section \ref{sec:intro}) is given by $\lambda^{-1} \int_{\lambda/2}^{\lambda} (ak)^2 \sin^2(kx) dx = (1/4)(ak)^2 $, we obtain:
\begin{equation}
    \Lambda = \frac{(ak)^2}{4\pi}\,\, \left(1 - c_p^+\,\sqrt{\Lambda} \right)^2  \,\, +\,\,\frac{1}{2}\,\, C_f(\Lambda^{-1/2} \Delta^+,z_0^u).
\label{lambda_simp}
\end{equation}
The SWARL-S model uses this value of $\Lambda$ in Equation \eqref{eq:z_0} to determine the roughness length using $\Delta = 3a$. Note that this model is expected to lose validity for wave steepness greater than about $ak \sim 0.28$ ($16^\circ$ degrees) because of the small angle approximation $\alpha \ll \pi$. 

\begin{figure}[h]
 \centerline{\includegraphics[width=37pc]{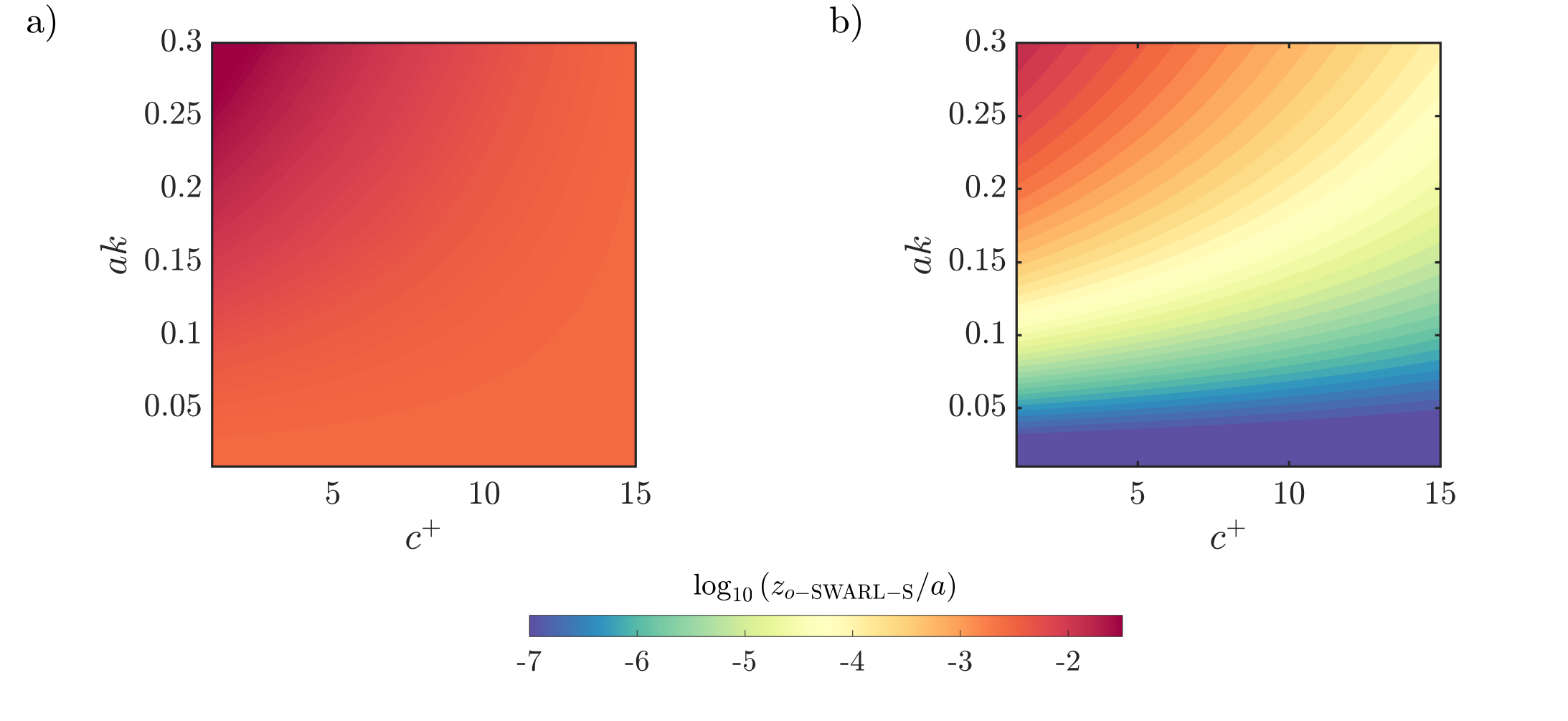}}
  \caption{Contours of predicted surface roughness length $z_0$ for monochromatic waves, using the simplified SWARL-S model (Equation~\ref{lambda_simp})  at  (a) $Re_\tau = 10^{2}$, and at (b) $Re_\tau = 10^{7}$.}
  \label{fig8}
\end{figure}

Using the SWARL-S model from Equation~\eqref{lambda_simp} we can perform a broad parametric study over a range of flow  ($Re_\tau$) and wave ($ak$ and $c^+$) parameters, without having to resort to numerical integration over wavesurface fields. Figure \ref{fig8} shows contour plots of the modeled surface roughness for a range of wave parameters at $Re_\tau = 10^2$ and $Re_\tau = 10^7$. The parametric study is done by assuming a smooth monochromatic wave with no stability corrections, that is to say $z_0^u \rightarrow 0$ and $\psi =0$. The results show that for the low $Re_\tau$, $z_0$ is proportionally a larger fraction of $a$ as compared to the high $Re_\tau$ cases, and illustrate that even in this simple case,  $z_0$ depends on multiple parameters, i.e., $ak$, $c^+$, and $Re_\tau$. Since the model is much simpler in practice than the full SWARL model, we also test its performance for predicting the surface roughness of multiscale wave fields, under the assumption that the peak wave represents the only contributor to drag. We approximate the full wave field   as a monochromatic   wave with parameters (amplitude, wave speed and wavelength) equal to those of the peak wave of a multiscale wave field. Again, in this way the SWARL-S model does not require numerical integration over the surface wave field realization, and instead only requires solving Equation \eqref{lambda_simp} via simple numerical root finding. 

\begin{figure}[h]
 \centerline{\includegraphics[width=37pc]{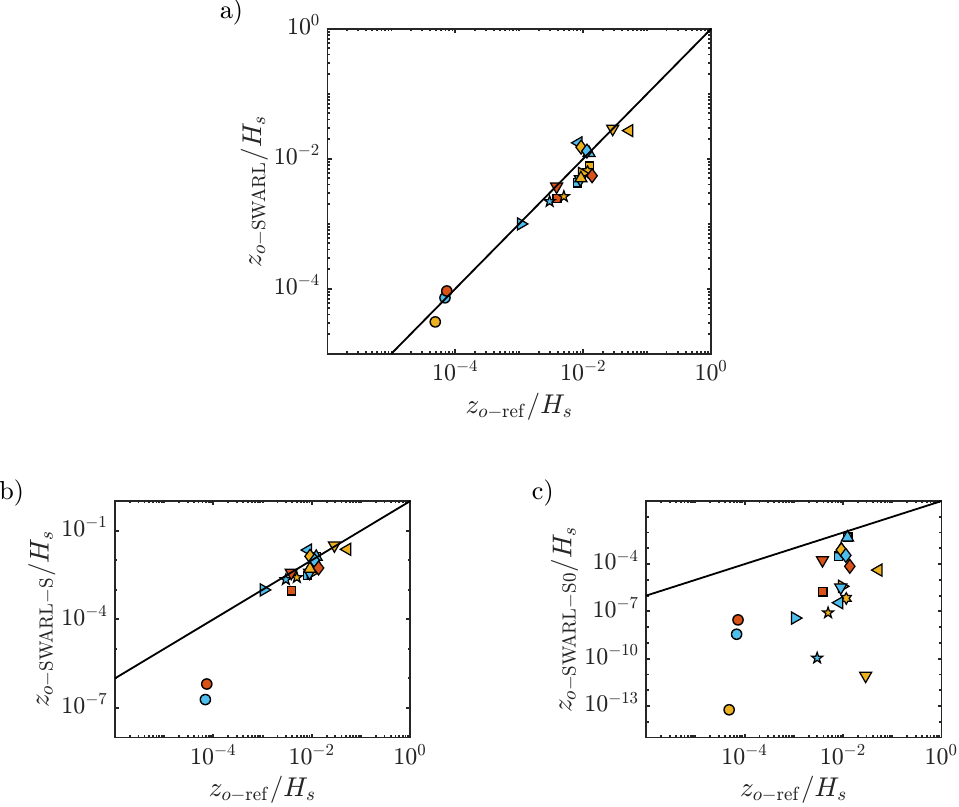}}
 \caption{Surface roughness predicted by a) SWARL, b) SWARL-S, and c) SWARL-S0 in the limit of $C_f=0$, versus values from reference data for all surface cases. For monochromatic wave cases, the surface roughness is normalized by the amplitude $a$. Symbols are as described in Table 1. The correlation and error parameters are $\rho=0.839$, $e_1= 0.178$ for SWARL, $\rho=0.745$, $e_1= 0.447$ for SWARL-S, and $\rho=0.062$, $e_1= 3.530$ for SWARL-S0, respectively.  }
  \label{fig9}
\end{figure}

Figure~\ref{fig9} compares roughness predictions from (a) the  full SWARL model  with (b) those from the simplified version, SWARL-S. The simplified model performs reasonably well for monochromatic wave cases, but consistently under predicts $z_0$ for  multiscale wave field cases. This under prediction likely arises because the SWARL-S model neglects the drag contribution of sub-peak waves, which, being slower, impart greater drag on the flow. We note in passing that in some cases  Equation \eqref{lambda_simp} yields no meaningful root. This scenario can occur when considering peak waves that are moving at high-speed at moderate Reynolds numbers, like the case of $c^+\approx 60$ from \cite{sullivan2014}. In selecting solutions to Equation \eqref{lambda_simp} we assumed that $u_\Delta^+ > c^+$ (i.e. that $\Delta$ falls above the critical layer). If we consider instead that $u_\Delta^+$ is smaller than the phase speed, then we must rearrange  Equation \eqref{lambda_simp} to consider the  negative root since $1-c^+\sqrt{\Lambda}<0$, leading to $-(\Lambda-C_f/2)^{1/2}  = (ak)/\sqrt{4\pi} ([1 – c^+ \sqrt{\Lambda}]$. This expression shows that if $C_f$ is large (i.e. low-Reynolds number and large drag from the viscous contribution), the air velocity near the wave is even smaller than the wave speed, allowing the equation to be satisfied. Since the SWARL approach is based on the assumption that $\Delta$ is at location above the critical layer where $u_\Delta^+ > c^+$, these cases are not included here. 

A further simplification of the SWARL-S model  (denoted SWARL-S0) can be formulated by considering the asymptotic limit $Re_\tau \rightarrow \infty$, for which the viscous contribution to total drag can be neglected ($C_f \rightarrow 0$). Under this assumption, Equation~\eqref{lambda_simp} reduces (again assuming $u^+ > c^+$) to a fully analytical form,
\begin{equation}
\Lambda =  \left[ \frac{ak}{ \sqrt{4\pi}+ak \, c^+ } \right]^2  ,
\label{lambda_simp_noCf}
\end{equation}
and
\begin{equation}
z_{o-\rm SWARL-S0} = 3a\, \exp \left(-\kappa \left[c^+ + \frac{\sqrt{4\pi}}{ak}  \right]  \right).
\label{lambda_simp_noCf}
\end{equation}
We note that while the exponential is reminiscent of the model by Kitaigorodskii \citep{Kitaigorodskii1970}, see discussion in \cite{rasex}, it is a different model and is not based on the assumption of a logarithmic profile in a moving frame of reference. 
The resulting predictions are shown in Fig.~\ref{fig9}c. In this further simplified case, roughness is   underpredicted even more broadly, including for several monochromatic wave cases. This additional underestimation is due to the neglect of finite-Reynolds-number effects: most monochromatic wave cases considered in our comparisons have $Re_\tau < 10^4$, where viscous contributions remain a significant proportion of the total drag. We conclude that although SWARL-S and SWARL-S0 are attractive for very idealized scenarios, they do not capture the complexity of realistic wind-wave interactions. In these cases, the full SWARL model including numerical integration over realizations of the wave field is required. 

\section{Conclusions}

This study develops the SWARL (Surface Wave Roughness Length) model, a physics-based framework for predicting the surface roughness length  of turbulent flow over wind-driven ocean waves. In contrast to traditional parameterizations that rely on empirical fits or highly simplified reduced representations (e.g., dependence only on wave age or wave steepness), the SWARL model requires no empirical tuning. It requires only numerical integrations over a representative realization of the wave surface height field at two consecutive times. This lack of tunable parameters makes it broadly applicable, while rooted in measurable physical quantities. By incorporating both pressure-drag mechanisms and wave-history effects—building on the approaches of  \citet{ayala2024} and \citet{meneveau2024}—the model captures the momentum lost by the airflow to the surface. We validate the SWARL model across 365 datapoints representing a variety of wind-wave conditions from DNS, WRLES, laboratory, and field experiments. When the wave surface can be statistically fully characterized, the SWARL model consistently outperforms widely used empirical models and algorithms in predicting both surface roughness and the drag coefficient, with significant improvements in correlation and error metrics. Even when surface characterization is approximate (e.g., cases where wave spectra are not available), the model remains competitive, providing predictions similar or slightly better than existing  approaches.  The generalizability of the  model  offers utility in a wide range of applications. For instance, it can enhance atmospheric boundary layer simulations in mesoscale and micro-mesoscale modeling frameworks, such as WRF-LES \citep{Deskos2021}. At larger scales, it can be used together with wave spectrum forecasting models (e.g. WaveWatch III) yielding $E(k,\theta)$, which can then be used to generate surface realizations to evaluate $z_0$ using SWARL.  Additionally, we envision that with {\it in-situ} field measurements of wave height and wind velocity, the proposed model could provide an efficient means to determine the surface momentum flux ($u_*^2$) from wave surface measurements, offering practical utility in interpreting field data. We also introduced a simplified version, SWARL-S, which is well-suited for idealized monochromatic waves but overpredicts drag for broadband or realistic sea states. A further simplified version, SWARL-S0, which—while analytically elegant—substantially underpredicts surface roughness due to the neglect of finite-Reynolds-number effects, reinforcing the need for the full SWARL model to capture realistic wind–wave interactions. Future extensions should focus on incorporating effects not addressed in the present framework, like swell conditions, wave breaking, and scalar (e.g., heat or moisture) fluxes, all challenges beyond the scope of this study.

\acknowledgments
We are grateful for fruitful conversations with Prof. Julie Lundquist.  This work was supported by the National Science Foundation and the Department of Energy (via NSF grant CBET-2401013) and by the U.S. Department of Energy, Office of Science Energy Earthshot Initiative, as part of the Addressing Challenges in Energy—Floating Wind in a Changing Climate (ACE-FWICC) Energy Earthshot Research Center.

\datastatement

Detailed explanations regarding the data used in this study can be found in supplementary material.

\bibliographystyle{ametsocV6}
\bibliography{jfm}

\end{document}